# Structural Properties, Impedance Spectroscopy and Dielectric Spin Relaxation of Ni-Zn Ferrite Synthesized by Double Sintering Technique

M. A. Ali[1], M. N. I. Khan[2], F.-U.-Z. Chowdhury[1], S. Akhter[2], and M. M. Uddin[1*]

[1]*Department of Physics, Chittagong University of Engineering and Technology (CUET), Chittagong-4349, Bangladesh.*
[2]*Materials Science Division, Atomic Energy Center, Dhaka-1000, Bangladesh.*

## Abstract

Structural properties, impedance, dielectric and electric modulus spectra have been used to investigate the sintering temperature ($T_s$) effect on the single phase cubic spinel $Ni_{0.6}Zn_{0.4}Fe_2O_4$ (NZFO) ceramics synthesized by standard ceramic technique. Enhancement of dielectric constants is observed with increasing $T_s$. The collective contribution of n-type and p-type carriers yields a clear peak in notable unusual dielectric behavior is successfully explained by the Rezlescu model. The non-Debye type long range dielectric relaxation phenomena is explained by electric modulus formalism. Fast response of the grain boundaries of the sample sintered at lower $T_s$ sample leading to small dielectric spin relaxation time, $\tau$ (several nanoseconds) have been determined using electric modulus spectra for the samples sintered at different $T_s$. Two clear semicircles in impedance Cole-Cole plot have also been successfully explained by employing two parallel RC equivalent circuits in series configuration taking into account no electrode contribution. Such a long relaxation time in NZFO ceramics could suitably be used in nanoscale spintronic devices.

*Keywords:* Ferrites; spin relaxation time; Dielectric relaxation; complex impedance spectroscopy; electric modulus.

## 1. Introduction

The spinel ceramics of $MFe_2O_4$ have attracted much attention in recent years for microwave applications owing to their remarkable magnetic properties and high resistivity values and low eddy current loss in high frequency operations. These properties make them promising candidates for the applications in radio frequency coils, transformer cores, rod antennas, magnetic cores of read write heads, high-speed digital tape or disk recording [1-5]. These magnetic insulators with high Neel temperature ($T_N$) can be used in spintronics and spin caloritronics as they naturally provide a non-zero magnetic moment along with spin dependent band gaps [6,7]. The spinel ferrites belong to $AB_2O_4$ structure such as Ni-Zn ferrites and having tetrahedral A site and octahedral B site in a unit cell are such potential candidates. The (Ni, Zn) $Fe_2O_4$ ferrite with a face-centered Fd$\bar{3}$m structure is one of the $MFe_2O_4$ phase family and could be described as $[Fe_{1-x}^{3+} Zn_x^{2+}]^A [Ni_{1-x}^{2+} Fe_{1+x}^{3+}]^B O_4$, where "A" represents the tetrahedral site, "B" represents the octahedral site, and "$x$" denotes the degree of inversion.

Nickel ions have preference for B-sites while zinc ions retain at A-sites and iron ions are distributed over both of the sites [5]. The Zn ions decrease during the calcination and sintering processes which creates cation vacancies and unsaturated oxygen ions. The excessive electrons on oxygen will bond with the neighboring $Fe^{3+}$ ions, which give rise to $Fe^{2+}$ ions. As a result, there is a possibility of electron hopping existing in *B*-sites within the ferrite leading to reduction of resistivity [8]. Additionally, the excellent relaxation behavior of Ni-Zn spinel ferrite makes it a strong candidate in the field of relaxors. The parameters which can sensibly alter the properties of ferrites are method of preparation, composition, sintering temperature, sintering time and rate of sintering [9,10]. Strategies will now need to be identified to probe and develop their properties in suitable conditions for technological applications. The structural and electrical properties of spinel ferrites are closely related. In some previous reports the structural, electrical and magnetic properties of Ni-Zn ferrites, prepared by different techniques, have been investigated [11-20].

The effects of sintering temperature on the structural, dielectric, resistivity and magnetic properties of (Ni,Zn)$Fe_2O_4$ ferrite have been investigated by Costa *et al.* [11], Kothawale *et al* [19], Rao *et al.* [20] and Hossain *et al.* [21]. Some authors have reported the impedance spectroscopy and dielectric relaxation as well as transport properties of (Ni,Zn)$Fe_2O_4$ ceramics [22,23]. But to the best of our knowledge, there are no reporting on the effect of sintering temperature on the impedance spectroscopy, non-Debye type dielectric relaxation, electron spin life time and electric modulus in conventional double sintering derived $Ni_{0.6}Zn_{0.4}Fe_2O_4$ ceramics. In order to take the advantages of this compound for eventual technological applications, an overall investigation is necessary.

In this work, we have reported the structural properties, impedance spectroscopy, dielectric spin relaxation and electric modulus of $Ni_{0.6}Zn_{0.4}Fe_2O_4$ ceramics over a wide frequency range, prepared by the conventional double sintering method. The contribution of grain and grain boundary resistance in impedance of the NZFO has been studied by impedance spectroscopy. The correlation between the dielectric spin relaxation time and sintering temperature, $T_s$ have been elucidated by analyzing the electric modulus with the established formalism.

## 2. Materials and Methods

The polycrystalline samples with basic chemical composition of $Ni_{0.6}Zn_{0.4}Fe_2O_4$ were synthesized by solid state reaction route using high purity (99.5%) (US Research Nanomaterials, Inc.) oxide precursors of nickel oxide (NiO), zinc oxide (ZnO) and iron oxide ($Fe_2O_3$) nano powders with particle sizes ~ 20-40, 15-35 and 35-45 nm, respectively in required stoichiometry. The analytical grade of powders was weighed according to the corresponding composition, and were mixed and ground for 3 hrs using an agate mortar and pestle. The slurry was dried and loosely pressed into cake using a hydraulic press. The cake is pre-sintered in air for 3 hrs at 900°C. The pre-sintered cake was removed from the furnace and it was crushed and ground again for 1 hr. The powder thus obtained was then pressed using a suitable die in the form of pellets of dimensions 8.4mm diameter and 1.1 mm thickness with a hydraulic press with a pressure of 10 kN for pellets using 5% polyvinyl alcohol solution as a binder. The samples were finally sintered at 1200, 1250 and 1300°C for 4 hrs in air at atmospheric pressure. The crystalline phases of the prepared samples were studied by X-ray diffraction (XRD) using Philips X'pert Pro X-ray diffractometer (PW3040) with Cu-K$_\alpha$ radiation ($\lambda$ = 1.5405 Å). The XRD data were collected at slow scan (2°/min) with a step size of 0.02° in a wide range of Bragg angles 15-70° at room temperature. Dielectric parameters such as capacitance, loss factor, impedance, and phase angles were measured at room temperature by a Wayne Kerr precision impedance analyzer (6500B) in the frequency range of 10 Hz to 100 MHz with drive voltage 0.5V. All the experimental data presented in this study were obtained at room temperature.

---

*Corresponding author: mohi@cuet.ac.bd

## 3. Results and Discussion

### 3.1. *Structural properties*

The XRD patterns of the samples sintered at different $T_s$ are shown in Fig. 1. Very sharp and well-defined peaks have been observed from the XRD pattern correspond to cubic spinel phase of the NZFO with Fd$\bar{3}$m space group symmetry. No secondary phase peaks have observed, indicating the high quality and single phase of synthesized sample. It is also observed that the diffraction peaks become narrower and sharper with the augment of $T_s$, representing an increase of the crystallite size, which has also been confirmed by SEM micrographs of the NZFO as shown in Fig. 1(b-d).The corresponding positions of all the sharp peaks are used to obtain the lattice parameters such as the lattice constant, $a_0$ (to determine accurate $a_0$, the extrapolation technique has been applied between the lattice parameter and the Nelson-Riley function $F(\theta)$, and the average value of $a_0$ is found to be 8.39187A°). The X-ray density, $\rho_x = 8M/N_A a_0^3$ ($N_A$ is Avogadro's number, $M$ is the molecular weight), bulk density, $\rho_b = M/V$ ($V$ is the volume of the samples and $V = \pi r^2 h$, $r$ = radius of the samples, $h$ = height of the samples), and the porosity $P(\%) = (((\rho_x - \rho_b)/\rho_x) \times 100)$ have also been calculated. The average values of $a_0$, $\rho_x$, $\rho_b$, and $P(\%)$ are found to be 8.39187Å, 5.33 gm/cm$^3$, 3.98 gm/cm$^3$ and 25.34%, respectively.

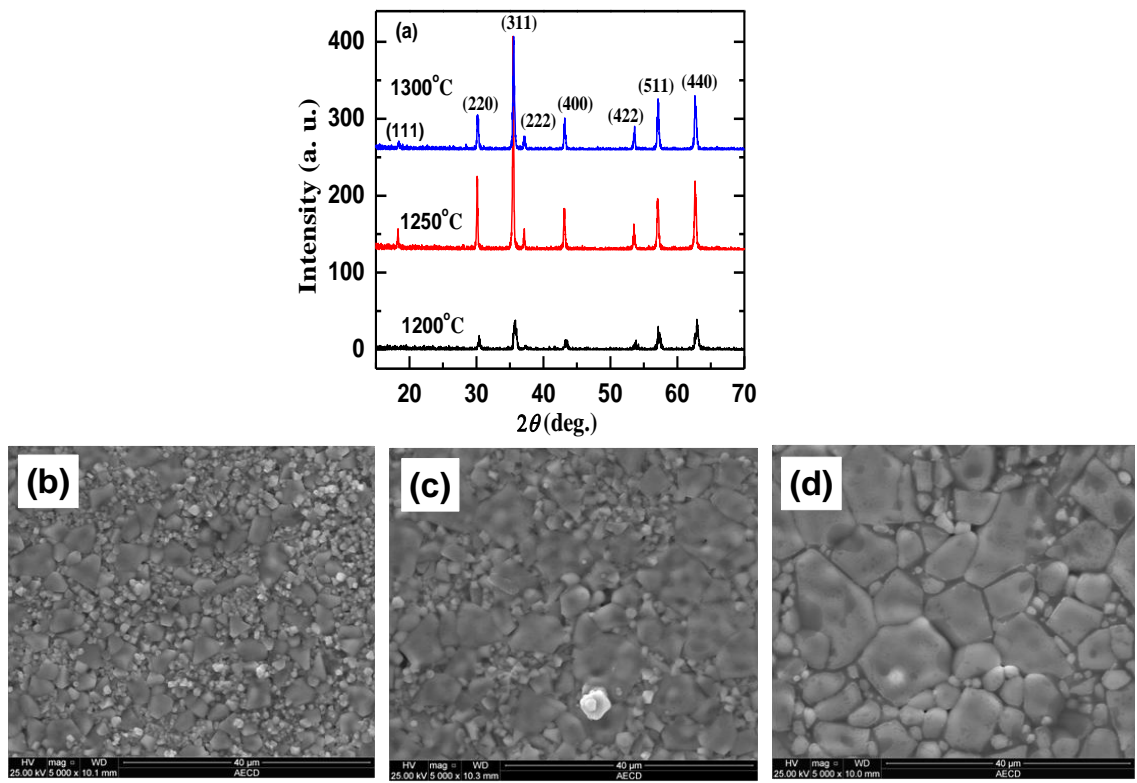

Fig. 1.(a) The XRD patterns of the Ni$_{0.6}$Zn$_{0.4}$Fe$_2$O$_4$(NZFO) ferrites at different sintering temperature ($T_s$). (b-d) show room temperature scanning electron microscopy (SEM) images of the NZFO sintered at $T_s$ 1200, 1250 and 1300°C, respectively.

A room temperature surface morphology or microstructure of the samples sintered at different $T_s$ has been studied using SEM micrograph as shown in Fig. 1 (b-d). The images exhibit clear grains and grain boundaries. The average grain size (grain diameter) has been determined from the micrographs by linear intercept technique. The values are found to be 1.4 - 7.8 $\mu$m, which are almost homogenously distributed throughout the sample surface. However, the grain size is obtained in the range of 6.2 - 7.8 $\mu$m for the sample sintered at 1300° C.

**3.2** *Dielectric relaxation properties*

The frequency dependence of the dielectric permittivity ($\varepsilon'$) [$\varepsilon' = CL/\varepsilon_0 A$ where, $C$ is the capacitance of the pellet, $L$ is the thickness of the pellet, $A$ is the cross-sectional area of the flat surface of the pellet and $\varepsilon_0$ is the constant of permittivity for free space] and tan$\delta$ of the NZFO ceramics have been measured at room temperature as depicted in Fig. 2. It is seen that the value of $\varepsilon'$ decreases with increase in frequency, which is typical characteristics of the polar dielectrics [24]. A remarkable increase in the value of $\varepsilon'$ has been observed for the sample sintered at higher temperature. Commonly, four types of polarizations (i.e., interfacial, dipolar, atomic and electronic) contribute in dielectric permittivity in ferrites [8, 22], among them dipolar and interfacial polarizations are strongly temperature dependent. Above mentioned polarization dynamics occur at the low frequency region, so permittivity increases faster leading to a high dielectric constant. The phonon and electron hopping enhance at measuring room temperature results the permittivity increases for the samples sintered at higher $T_s$ (Fig. 2). It is evident that Zn is evaporated at higher sintering temperature and creates vacancies in the crystal. The Fe elements have two valence states of $Fe^{2+}$ and $Fe^{3+}$. Some of the ions $Fe^{3+}$ from the *B* site convert to $Fe^{2+}$ ions due to compensate the Zn loss which accelerates the hopping process between the Fe ions in +2 and +3 states. More and more charges reach the grain boundary due to increase in polarization process results accumulation of charges at the grain boundary of the NZFO ceramics upturns. As a consequence, the permittivity increases for the samples sintered at higher Ts. Moreover, at the lower frequency region increase of permittivity is more prominent. The decrease of permittivity with the increase of frequency could be understood by Maxwell-Wagner type dielectric structure. The development of bigger grains and boundaries as well as boundary reaction become more remarkable that the dipoles and boundary ions cannot be able to follow changes of the field causing faster polarizationin high frequencies region leading to decrease even zero value of $\varepsilon'$.

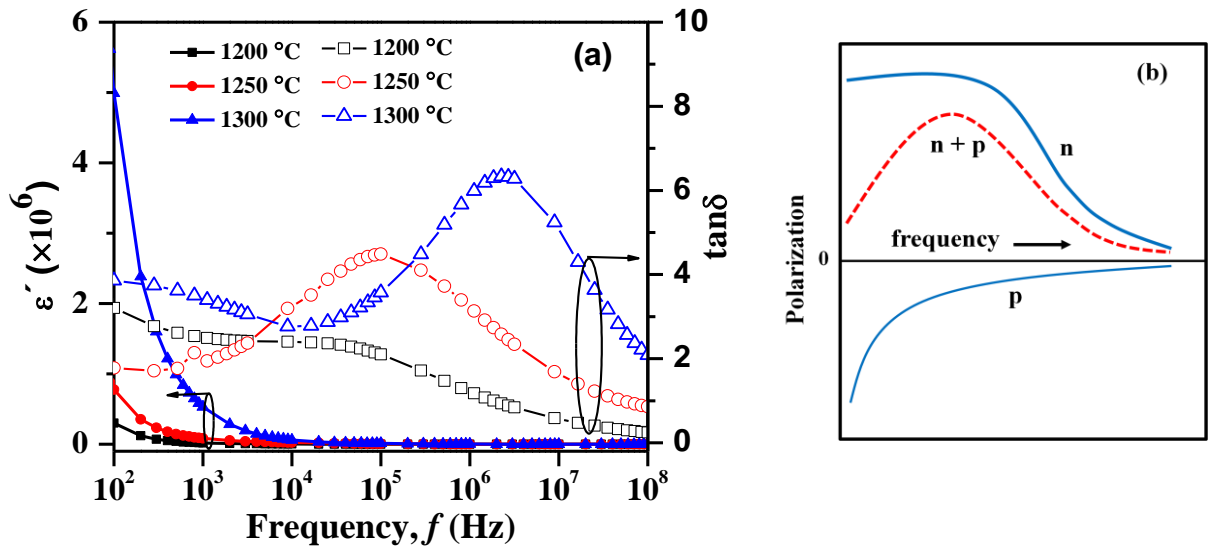

Fig. 2. Variation of the dielectric permittivity ($\varepsilon'$) and loss factor (tan$\delta$) of the $Ni_{0.6}Zn_{0.4}Fe_2O_4$ ferrites sintered at different $T_s$ as a function of frequency. (b) shows the Rezlescu model for the dielectric relaxation peaks.

Frequency dependent loss factor (tan$\delta$, is defined as $\varepsilon''/\varepsilon'$) for the samples sintered at different $T_s$ for the NZFO is shown in Fig. 2. The curve shows the dielectric relaxation processes (peaks) at the particular frequency which are shifted at the higher frequency value. The dielectric relaxation peaks appear when the externally applied AC electric field becomes equal to that of the jumping frequency of localized electric charge carrier [25]. It is reported that a visible peak in tan$\delta$ with increasing angular frequency could be observed if the value of $\varepsilon'$ decreases and imaginary dielectric constant, $\varepsilon''$, slowly decreases with frequency [26, 27]. The obtained unusual behavior of dielectric properties in the NZFO could be explained by the Rezlescu model as shown in Fig. 2(b) [28]. It states that the contributions of n-type and p-type carriers decrease very slowly and rapidly with angular frequency increases, respectively. However, the collective contribution of both types of carriers to the dielectric polarization yields a clear peak with the frequency. The observed dielectric peak of the NZFO is closely compared with the Rezlescu model (red curve in Fig. 2 (b)). The electric conduction in ferrites arises due to the electron exchange between $Fe^{2+}$ and $Fe^{3+}$ and hole transfer between $Ni^{3+}$ and $Ni^{2+}$ [29] at the octahedral (*B*) sites which is similar to that of dielectric polarization in ferrites [28]. The Fe and Ni ions are produced by the following mechanism; $Ni^{2+} + Fe^{3+} \leftrightarrow Ni^{3+} + Fe^{2+}$, $Fe^{3+} \leftrightarrow Fe^{2+} + e^-$.

## 3.3 Electric modulus

Complex electric modulus is a well-known and powerful tool which has been effectively used to analyze the electrical response of the materials such as nature of polycrystalline in samples (homogenous or inhomogeneous) and the electrical relaxation in electronically and ionically conducting materials as a microscopic property of the materials [30]. The real part of electric modulus ($M'$) and impedance ($Z'$) as a function of frequency for the samples sintered at different $T_s$ has been depicted in Fig. 3. The value of $M'$ reaches a maximum at high frequency region with reaching to zero at low frequency indicating the electrode polarization contribution is negligible [31]. The value of $Z'$ decreases monotonically with increases in frequency as seen in Fig. 3. It is also observed that the value of $Z'$ shows dispersion pattern at different sintering temperature, $T_s$ in low frequency region followed by a plateau. Finally all the curves coalesce with approaching zero value representing $Z'$ is independent of frequency. The space charge has lesser time to relax at higher frequency region and the recombination is faster which reduces the space charge polarization leading to a merge of the curves of all the samples sintered at different $T_s$ [32].

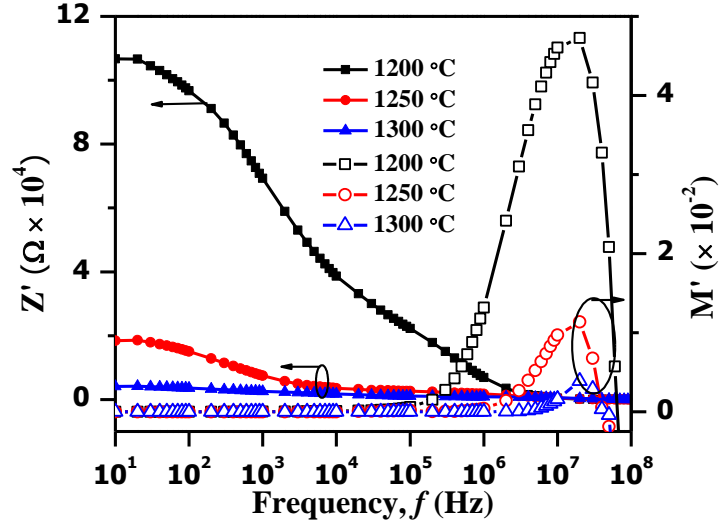

Fig.3. Frequency dependent impedance real part ($Z'$) and electric modulus ($M'$) of the $Ni_{0.6}Zn_{0.4}Fe_2O_4$ ferrites sintered at different $T_s$.

A combined plot of imaginary electric modulus $M''$ [the complex electric modulus $M^* = M' + jM''$, $M' = \varepsilon'/(\varepsilon'^2 + \varepsilon''^2)$ and $M'' = \varepsilon''/(\varepsilon'^2 + \varepsilon''^2)$] and impedance $Z''$ as a function of frequency is useful to detect the effect of the smallest capacitance, the largest resistance and assists to distinguish whether a relaxation process is due to short or long-range movement of charge carriers as proposed by Sinclair and West [33]. They also reported that if the peak in $M''$ versus frequency and $Z''$ versus frequency will occur at the same frequency then the process is long range and if these peaks will occur at different frequencies then the process is localized. The plot $M''$ versus $Z''$ as a function of frequency for the samples sintered at different $T_s$ is shown in Fig. 4 (a). The presence of localized movement of charge carriers and departure from ideal

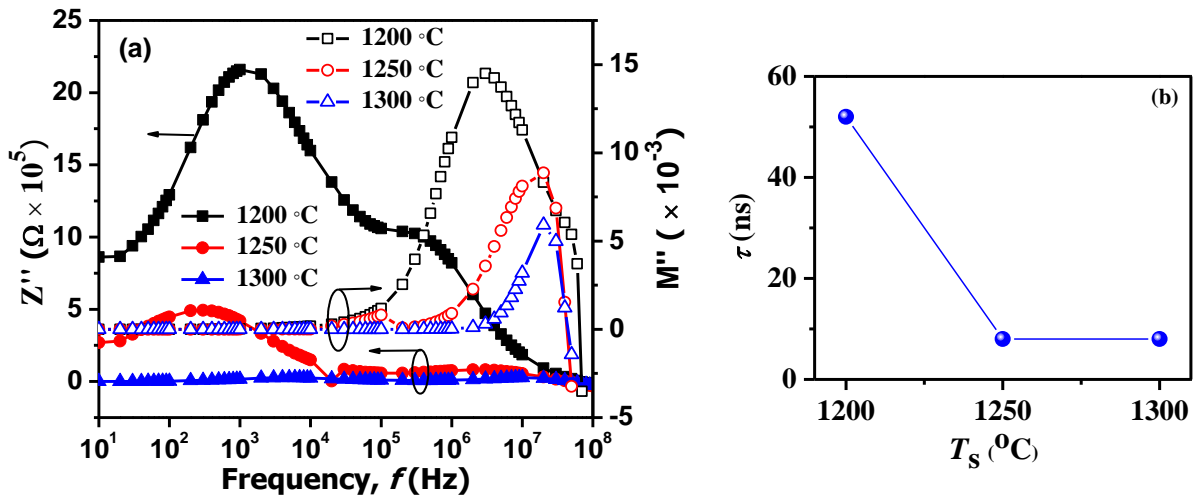

Fig. 4. (a) Frequency dependence of imaginary part of impedance ($Z''$) and electric modulus ($M''$) of the $Ni_{0.6}Zn_{0.4}Fe_2O_4$ ferrites sintered at various $T_s$. (b) The dielectric spin relaxation time for the samples sintered at different $T_s$.

Debye-like behavior has been proved by the separation between these peaks as shown in Fig. 4 (a) [34]. Moreover, the relaxation of the electric field in the materials is denoted by the electric modulus since the electric displacement remains constant. Thus, the peak value in $M''$ represents the real dielectric spin relaxation process of the materials where the peak frequency is indicative of transition from long to short range mobility with increase in frequency. The relaxation time is determined by the relation $\tau_{M''}= 1/(2\pi f_{M''})$, $f_{M''}$ is the frequency at which $M''(\omega)$ is maximum ($M''_{max}$). The dielectric spin relaxation time ($\tau$) has been calculated from $M''$ vs. frequency plot as shown in Fig. 4 (b). Particular noteworthy is that the value of $\tau$ is in the range of several nano seconds and decreases with increasing sintering temperarure, $T_s$. At higher sintering temperarure, the grain boundary movement faster than that of low sintering temperarure results the dielectric spin easily relax. However, the spin-relaxation times of semiconductor materials are typically ~100 ns, 150 ns in cobalt nanoparticle and they are on the order of picoseconds in bulk metals due to the high density of scattering centres [35, 36]. Such a long dielectric spin-relaxation time in the NZFO ceramics can be very useful in nanoscale spintronic devices. The imaginary part of $Z''$ shows monotonous decrease with increasing frequency of the samples sintered at various $T_s$ as illustrated in Fig. 4 (a).

### 3.4 *Impedance spectroscopy*

The complex impedance spectrum (Nyquist plots/ Cole-Cole plot) technique is a well-known and powerful tool which has been effectively used for complete understands the electrical properties of the electro-ceramic materials such as impedance of electrodes, grain and grain boundaries. It also provides the information about the resistive (real part) and reactive (imaginary part) components of a material. A Cole-Cole plot typically comprises of two successive semicircles: first semicircle is due to the contribution of the grain boundary at low frequency and second is due to the grain or bulk properties at high frequency of the materials.

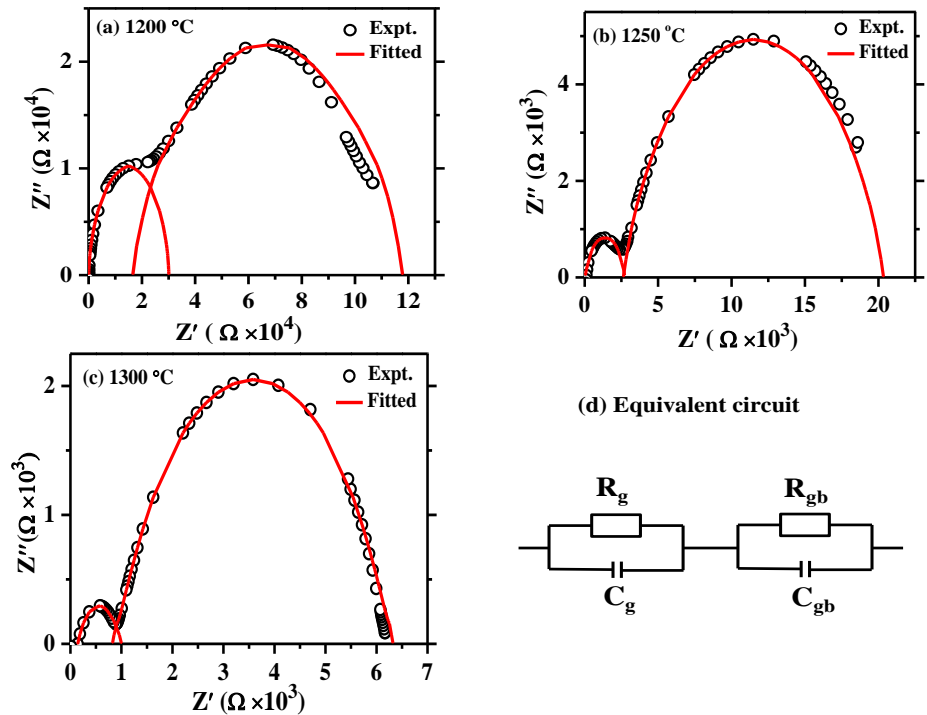

Fig. 5.(a-c) Cole–Cole plot of complex impedance of the $Ni_{0.6}Zn_{0.4}Fe_2O_4$ ferrites sintered at different $T_s$. (d) shows the equivalent circuit.

Fig. 5 shows the room temperature complex impedance or Cole–Cole plot for the samples sintered at different $T_s$ as a function of frequency. Two partially overlapping semi-circles are found whose centers are lying below the real axis, indicating a non-Debye type of relaxation process as discussed in previous section. A sample is assumed that a microstructure made up of parallel conducting plates (grains) separated by resistive plates (grain boundaries). According to the Maxwell–Wagner two layer model, the phenomena is typically related to the existence of a distribution of relaxation time, in which the resulting complex impedance is composed of two overlapping semicircles [37, 38].These two semicircles could be modeled by two parallel RC equivalent circuits in series configuration as shown in Fig. 5 (d).The complex impedance of a system at an applied frequency can be written as sum of the real and imaginary parts: $Z^*(\omega) = Z'(\omega) + iZ''(\omega)$, where $Z'$ and $Z''$ of the impedance can be represented as,

$$Z' = \frac{R_g}{\left(1+\omega_g C_g R_g\right)^2} + \frac{R_{gb}}{\left(1+\omega_{gb} C_{gb} R_{gb}\right)^2} \qquad (1)$$

$$Z'' = \frac{R_g^2}{1+\left(\omega_g C_g R_g\right)^2} + \frac{R_{gb}^2}{1+\left(\omega_{gb} C_{gb} R_{gb}\right)^2} \quad (2)$$

Where $R_g$ and $C_g$ represent the resistance and capacitance of the grain and $R_{gb}$ and $C_{gb}$ represent the corresponding terms for grain boundary, while $\omega_g$ and $\omega_{gb}$ are the frequencies at the peaks of the semicircles for grain and grain boundary, respectively. The resistances are calculated from the circular arc intercepts on the $Z$ axis, while the capacitances are derived from the maximum height of the circular arcs. The maximum height in each semicircle is $Z = -Z''$. We fit our data using the above equations and obtained parameters are listed in Table-1. It is noticed that the circular arcs in the low frequency regions are not completed which may be due to the fact that grain boundary resistance is out of measurement scale.

Table 1: Impedance parameters of the $Ni_{0.6}Zn_{0.4}Fe_2O_4$ ferrites at room temperature.

| $T_S$ (°C)  Parameters | 1200 | 1250 | 1300 |
|---|---|---|---|
| $\tau_M$(ns) | 50 | 8 | 8 |
| $R_g$ (kΩ) | 29.9 | 2.66 | 0.93 |
| $R_{gb}$ (kΩ) | 101.2 | 17.7 | 5.3 |
| $C_g$ (nF) | 0.18 | 0.03 | 0.017 |
| $C_{gb}$ (nF) | 1.57 | 29.9 | 7.5 |
| $\tau_g$ ($\tau_g=1/\omega_g=R_g C_g$) (ns) | 538 | 80 | 16 |
| $\tau_{gb}$ ($\tau_{gb}=1/\omega_{gb}=R_{gb} C_{gb}$) (μs) | 158 | 529 | 39.8 |

## 4. Conclusions

In summary, we have prepared and studied the structural, dielectric and electrical properties of single phase cubic spinel $Ni_{0.6}Zn_{0.4}Fe_2O_4$ ferrites at various sintering temperature. An increase in dielectric constants is explained by the electron hopping mechanism between $Fe^{2+}$ and $Fe^{3+}$ states. Remarkable unusual dielectric behavior is due to the collective contribution of both n and p-type carriers which is fruitfully explained by Rezlescu model. Irregular long range and non-Debye type dielectric relaxation are attributed by the electric modulus. The complex impedance spectra show contribution of both grain and grain boundary effects in the electrical properties, endorsed by the classical Cole-Cole plot. Long dielectric spin relaxation times in nanosecond range have also been determined from the electric modulus spectra that make the NZFO as a promising candidate for future nanoscale spintronic devices.

**Acknowledgements:** The authors are grateful to the authority of CUET (DRE grant) for providing financial support for this work.